\DeclareMathAlphabet{\pazocal}{OMS}{zplm}{m}{n}
\newcommand{\Jpsi}{J\!/\!\psi}

\documentclass[manyauthors,nocleardouble,COMPASS]{cernphprep}
\usepackage{bm}
\usepackage{cite}  
\usepackage{lineno}  
\usepackage{booktabs}
\usepackage{multirow}
\usepackage{color}
\usepackage{amsmath}
\usepackage[figuresright]{rotating}

\newcommand{\GeV}{\mathrm{GeV}}
\usepackage{calrsfs}
\usepackage{hyperref}
\usepackage[T1]{fontenc}

\hypersetup{
    colorlinks,
    citecolor=blue,
    filecolor=blue,
    linkcolor=blue,
    urlcolor=blue
}

\graphicspath{ {figures/}  {figures-new/}  }

\begin{document}


\newcommand\PRLSTYLE{no}

\begin{titlepage}
\PHnumber{2017--165}
\PHdate{\today}

\title{	Search for muoproduction of $X(3872)$ at COMPASS and indication  of a new state $\widetilde{X}(3872)$}


\Collaboration{The COMPASS Collaboration}
\ShortAuthor{The COMPASS Collaboration}
\ShortTitle{$X(3872)$ photoproduction at COMPASS}

\begin{abstract} 
We have searched for exclusive production of exotic charmonia in the reaction  \\
\mbox{$\mu^+~N \rightarrow \mu^+ (\Jpsi\pi^+\pi^-)\pi^{\pm}~N'$} using COMPASS data collected with incoming 
muons of 160~GeV/$c$  and 200~GeV/$c$ momentum. In the $\Jpsi\pi^+\pi^-$ mass distribution we observe a signal with a statistical significance of 4.1 $\sigma$. Its mass and width are consistent with those of the $X(3872)$. The shape of the $\pi^+\pi^-$ mass distribution from the observed decay into $\Jpsi\pi^+\pi^-$ shows disagreement with previous observations for $X(3872)$.
The observed signal  may be interpreted as a possible evidence of a new charmonium state. It could be associated with a neutral partner of $X(3872)$ with $C = -1$ predicted by a tetraquark model.
 The product of cross section and  branching fraction of the decay of the observed state
 into $\Jpsi\pi^+\pi^-$ is determined to be 71$\pm$28(stat)$\pm$39(syst)~pb.
\end{abstract}

\vspace*{60pt}
\begin{flushleft}

\end{flushleft}

\vfill
\Submitted{(submitted to PLB)}
\end{titlepage}

{\pagestyle{empty}
\label{app:collab}
\renewcommand\labelenumi{\textsuperscript{\theenumi}~}
\renewcommand\theenumi{\arabic{enumi}}
\begin{flushleft}
%
%
\section*{The COMPASS Collaboration}
\label{app:collab}
\renewcommand\labelenumi{\textsuperscript{\theenumi}~}
\renewcommand\theenumi{\arabic{enumi}}
\begin{flushleft}
M.~Aghasyan\Irefn{triest_i},
R.~Akhunzyanov\Irefn{dubna}, 
M.G.~Alexeev\Irefn{turin_u},
G.D.~Alexeev\Irefn{dubna}, 
A.~Amoroso\Irefnn{turin_u}{turin_i},
V.~Andrieux\Irefnn{illinois}{saclay},
N.V.~Anfimov\Irefn{dubna}, 
V.~Anosov\Irefn{dubna}, 
A.~Antoshkin\Irefn{dubna}, 
K.~Augsten\Irefnn{dubna}{praguectu}
% ...
\end{flushleft}
%
%
\begin{Authlist}
\item \Idef{aveiro}{University of Aveiro, Dept.\ of Physics, 3810-193 Aveiro, Portugal}
\item \Idef{bochum}{Universit\"at Bochum, Institut f\"ur Experimentalphysik, 44780 Bochum, Germany\Arefs{l}\Aref{s}}
\item \Idef{bonniskp}{Universit\"at Bonn, Helmholtz-Institut f\"ur  Strahlen- und Kernphysik, 53115 Bonn, Germany\Arefs{l}}
\item \Idef{bonnpi}{Universit\"at Bonn, Physikalisches Institut, 53115 Bonn, Germany\Arefs{l}}
\item \Idef{brno}{Institute of Scientific Instruments, AS CR, 61264 Brno, Czech Republic\Arefs{m}}
\item \Idef{calcutta}{Matrivani Institute of Experimental Research \& Education, Calcutta-700 030, India\Arefs{n}}
\item \Idef{dubna}{Joint Institute for Nuclear Research, 141980 Dubna, Moscow region, Russia\Arefs{o}}
\item \Idef{erlangen}{Universit\"at Erlangen--N\"urnberg, Physikalisches Institut, 91054 Erlangen, Germany\Arefs{l}}
\item \Idef{freiburg}{Universit\"at Freiburg, Physikalisches Institut, 79104 Freiburg, Germany\Arefs{l}\Aref{s}}
\item \Idef{cern}{CERN, 1211 Geneva 23, Switzerland}
\item \Idef{liberec}{Technical University in Liberec, 46117 Liberec, Czech Republic\Arefs{m}}
\item \Idef{lisbon}{LIP, 1649-003 Lisbon, Portugal\Arefs{p}}
\item \Idef{mainz}{Universit\"at Mainz, Institut f\"ur Kernphysik, 55099 Mainz, Germany\Arefs{l}}
\item \Idef{miyazaki}{University of Miyazaki, Miyazaki 889-2192, Japan\Arefs{q}}
\item \Idef{moscowlpi}{Lebedev Physical Institute, 119991 Moscow, Russia}
\item \Idef{munichtu}{Technische Universit\"at M\"unchen, Physik Dept., 85748 Garching, Germany\Arefs{l}\Aref{r}}
\item \Idef{nagoya}{Nagoya University, 464 Nagoya, Japan\Arefs{q}}
\item \Idef{praguecu}{Charles University in Prague, Faculty of Mathematics and Physics, 18000 Prague, Czech Republic\Arefs{m}}
\item \Idef{praguectu}{Czech Technical University in Prague, 16636 Prague, Czech Republic\Arefs{m}}
> 
\item \Idef{protvino}{NRC {\guillemotleft}Kurchatov Institute{\guillemotright} -- IHEP, 142281 Protvino, Russia}
\item \Idef{saclay}{IRFU, CEA, Universit\'e Paris-Saclay, 91191 Gif-sur-Yvette, France\Arefs{s}}
\item \Idef{taipei}{Academia Sinica, Institute of Physics, Taipei 11529, Taiwan\Arefs{tw}}
\item \Idef{telaviv}{Tel Aviv University, School of Physics and Astronomy, 69978 Tel Aviv, Israel\Arefs{t}}
\item \Idef{triest_u}{University of Trieste, Dept.\ of Physics, 34127 Trieste, Italy}
\item \Idef{triest_i}{Trieste Section of INFN, 34127 Trieste, Italy}
\item \Idef{turin_u}{University of Turin, Dept.\ of Physics, 10125 Turin, Italy}
\item \Idef{turin_i}{Torino Section of INFN, 10125 Turin, Italy}
\item \Idef{tomsk}{Tomsk Polytechnic University,634050 Tomsk, Russia\Arefs{nauka}}
\item \Idef{illinois}{University of Illinois at Urbana-Champaign, Dept.\ of Physics, Urbana, IL 61801-3080, USA\Arefs{nsf}}
\item \Idef{warsaw}{National Centre for Nuclear Research, 00-681 Warsaw, Poland\Arefs{u}}
\item \Idef{warsawu}{University of Warsaw, Faculty of Physics, 02-093 Warsaw, Poland\Arefs{u}}
\item \Idef{warsawtu}{Warsaw University of Technology, Institute of Radioelectronics, 00-665 Warsaw, Poland\Arefs{u} }
\item \Idef{yamagata}{Yamagata University, Yamagata 992-8510, Japan\Arefs{q} }
\end{Authlist}
%
%
\renewcommand\theenumi{\alph{enumi}}
\begin{Authlist}
\item [{\makebox[2mm][l]{\textsuperscript{\#}}}] Corresponding authors
\item [{\makebox[2mm][l]{\textsuperscript{*}}}] Deceased
\item \Adef{a}{Also at Instituto Superior T\'ecnico, Universidade de Lisboa, Lisbon, Portugal}
\item \Adef{b}{Also at Dept.\ of Physics, Pusan National University, Busan 609-735, Republic of Korea and at Physics Dept., Brookhaven National Laboratory, Upton, NY 11973, USA}
\item \Adef{ictp}{Also at Abdus Salam ICTP, 34151 Trieste, Italy}
\item \Adef{r}{Supported by the DFG cluster of excellence `Origin and Structure of the Universe' (www.universe-cluster.de) (Germany)}
\item \Adef{p2i}{Supported by the Laboratoire d'excellence P2IO (France)}
\item \Adef{d}{Also at Chubu University, Kasugai, Aichi 487-8501, Japan\Arefs{q}}
\item \Adef{x}{Also at Dept.\ of Physics, National Central University, 300 Jhongda Road, Jhongli 32001, Taiwan}
\item \Adef{e}{Also at KEK, 1-1 Oho, Tsukuba, Ibaraki 305-0801, Japan}
\item \Adef{g}{Also at Moscow Institute of Physics and Technology, Moscow Region, 141700, Russia}
\item \Adef{h}{Present address: RWTH Aachen University, III.\ Physikalisches Institut, 52056 Aachen, Germany}
\item \Adef{y}{Also at Dept.\ of Physics, National Kaohsiung Normal University, Kaohsiung County 824, Taiwan}
\item \Adef{regen}{Also at Institut f\"ur Theoretische Physik, Universit\"at T\"ubingen, 72076 T\"ubingen, Germany}
\item \Adef{infn}{Also at \ref{LItriest_i})}
\item \Adef{turin_p}{Also at University of Eastern Piedmont, 15100 Alessandria, Italy}
\item \Adef{i}{Present address: Uppsala University, Box 516, 75120 Uppsala, Sweden}
\item \Adef{c}{    Supported by the DFG Research Training Group Programmes 1102 and 2044 (Germany)} 
%
%
\item \Adef{l}{    Supported by BMBF - Bundesministerium f\"ur Bildung und Forschung (Germany)}
\item \Adef{s}{    Supported by FP7, HadronPhysics3, Grant 283286 (European Union)}
\item \Adef{m}{    Supported by MEYS, Grant LG13031 (Czech Republic)}
\item \Adef{n}{    Supported by SAIL (CSR) and B.Sen fund (India)}
\item \Adef{o}{    Supported by CERN-RFBR Grant 12-02-91500}
\item \Adef{p}{\raggedright 
                   Supported by FCT - Funda\c{c}\~{a}o para a Ci\^{e}ncia e Tecnologia, COMPETE and QREN, Grants CERN/FP 116376/2010, 123600/2011 
                   and CERN/FIS-NUC/0017/2015 (Portugal)}
\item \Adef{q}{    Supported by MEXT and JSPS, Grants 18002006, 20540299, 18540281 and 26247032, the Daiko and Yamada Foundations (Japan)}
\item \Adef{tw}{   Supported by the Ministry of Science and Technology (Taiwan)}
\item \Adef{t}{    Supported by the Israel Academy of Sciences and Humanities (Israel)}
\item \Adef{nauka}{Supported by the Russian Federation  program ``Nauka'' (Contract No. 0.1764.GZB.2017) (Russia)}
\item \Adef{nsf}{  Supported by the National Science Foundation, Grant no. PHY-1506416 (USA)}
\item \Adef{u}{    Supported by NCN, Grant 2017/26/M/ST2/00498 (Poland)}
\end{Authlist}

\end{flushleft}

\clearpage
}

\setcounter{page}{1}

The exotic hadron $X(3872)$ was first discovered in 2003 by the Belle collaboration \cite{Choi:2003ue}  and constitutes the first in a long
 series of new charmonium-like hadrons at masses above 3.8~GeV/$c^2$. 
The $X(3872)$ was observed as a narrow peak in the $\Jpsi\pi^+\pi^-$ mass spectrum originating from the decay $B^{\pm}\rightarrow  
K^{\pm}\Jpsi\pi^+\pi^-$.  
Subsequently,  this state has also been observed in numerous reaction channels and final states: in $e^+e^-$ collisions by Belle 
\cite{Gokhroo:2006bt,Adachi:2008sua,Bhardwaj:2011dj,Choi:2011fc}, Babar \cite{Aubert:2004ns,Aubert:2005zh,Aubert:2006aj,
Aubert:2007rva,Aubert:2008gu,Aubert:2008ae,delAmoSanchez:2010jr} and BESIII \cite{Ablikim:2013dyn} and in hadronic 
interactions by CDF 
\cite{Acosta:2003zx,Abulencia:2005zc,Abulencia:2006ma,Aaltonen:2009vj}, D0 \cite{Abazov:2004kp}, LHCb \cite{Aaij:2011sn,
Aaij:2013zoa,Aaij:2014ala},  ATLAS \cite{Aaboud:2016vzw} and CMS \cite{Chatrchyan:2013cld}.
 The current world average for the mass of the $X(3872)$ is 3871.69$\pm$0.17 MeV/$c^2$ \cite{pdg}, which is very close to the 
 $D^0\bar{D}^{*0}$ threshold at 3871.81$\pm$0.09 MeV/$c^2$. However, the decay width of this state was not determined yet
 as in all experiments the measured widths were compatible with the experimental resolution. Thus only an upper limit for the natural width 
 $\Gamma_{X(3872)}$  of about 1.2 MeV/$c^2$ (CL $ = 90$\%)  exists \cite{Choi:2011fc}.
 The spin, parity and charge-conjugation quantum numbers $J^{PC}$ of the $X(3872)$ were determined by LHCb to be 1$^{++}$ 
 \cite{Aaij:2013zoa,LHCb-dipion}. 
  Charged partners of the $X(3872)$ have not been observed \cite{Aubert:2004zr}.
 The $X(3872)$ hadron is peculiar in several aspects and its nature is still not well understood. In 
 particular, approximately equal probabilities to decay into $\Jpsi$3$\pi$ and $\Jpsi$2$\pi$ final states $\pazocal{B}(X(3872)\rightarrow 
 \Jpsi\omega) / \pazocal{B}(X(3872)\rightarrow \Jpsi\pi^+\pi^-)=0.8\pm0.3$ \cite{Gparity} indicate large isospin-symmetry breaking. There 
 are several interpretations of this hadron: pure $c\bar{c}$-state, tetraquark, meson-meson molecule, $c\bar{c}g$ meson, glueball, or 
 others (see reviews \cite{Lebed:2016hpi, Hosaka:2016pey, Chen:2016qju}). In addition to knowing mass and quantum numbers of this 
 state, the measurement of its width would provide a crucial input to narrow down speculations on its nature. Currently such a 
 measurement can only be done by performing energy scans in $p\bar{p}$ annihilations, as it is foreseen at FAIR \cite{FAIR, FAIR2}. 
 
In this Letter, we report on a search for $X(3872)$  produced by virtual photons in the charge-exchange reaction
\begin{equation}
\label{reak1}
\gamma^* N \rightarrow X^{0}\pi^{\pm}N' 
\end{equation}
at COMPASS. Here, $N$ denotes the target nucleon, $N'$ the unobserved recoil system and $X^0$ an intermediate state decaying into $\Jpsi\pi^{+}\pi^{-}$. The possibility to observe the production of 
$X(3872)$ in this reaction was first mentioned in Ref. \cite{BingAnLi}.

The COMPASS experiment \cite{Abbon:2007pq} is situated at the M2 beam line of the CERN Super Proton Synchrotron. 
The data used in the present analysis were obtained by scattering positive muons
 of 160~GeV/$c$ or 200~GeV/$c$ momentum off solid $^6$LiD or NH$_3$ targets. The total data set 
 accumulated between 2003 and 2011 was used.
The target material was arranged in two or three cylindrical cells placed along the
 beam direction. It was longitudinally or transversely polarized with respect to this direction. 
The polarization is opposite in consecutive target cells, and it is reversed periodically during data taking. After combining data with 
opposite polarization, possible effects from residual target polarization have negligible influence on this analysis.
Particle tracking and identification were performed  using a two-stage spectrometer, covering a wide momentum range from about 1~GeV/$c$ up to the beam 
momentum. The event trigger was based on scintillator hodoscopes and hadron calorimeters. Different trigger 
schemes were used for the different data sets. Possible differences in trigger efficiencies are expected to cancel in the determination of 
absolute production rates, which are obtained using a normalization process that was recorded in parallel, see below.
Beam halo muons were rejected by veto counters located upstream of the target. 
 \ifthenelse{\equal{\PRLSTYLE}{yes}}{\begin{figure}}{\begin{figure}[h]}
 \begin{center}
    \includegraphics[width=220px]{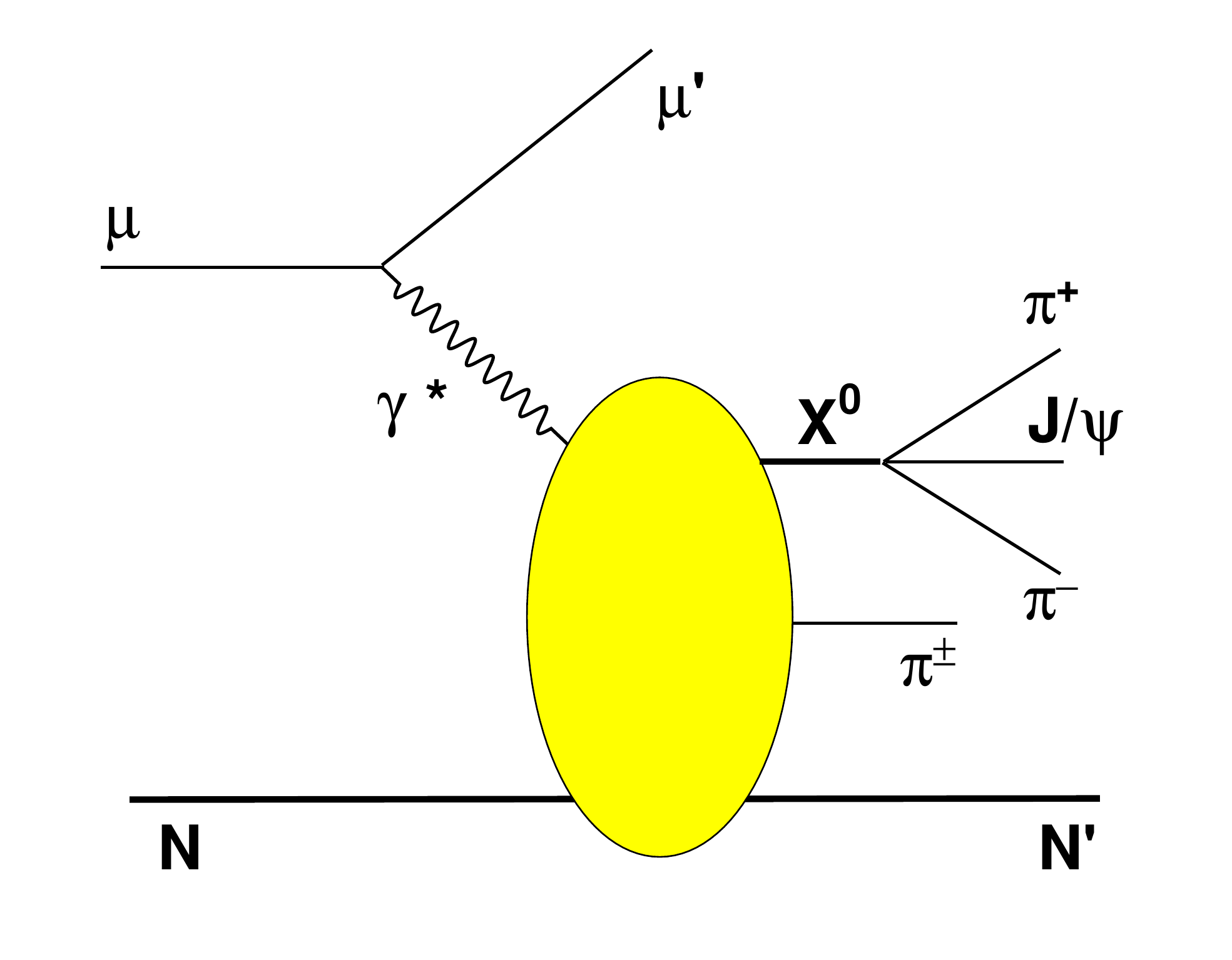}
   \end{center}
     \caption{\label{fig:diag0}
      Diagram for exclusive muoproduction of $X^0\pi^{\pm}$ in reaction (\ref{reaction3}).
  }
\end{figure}

 The main subject of this Letter is the study of muoproduction of an $X^0$ 
 in the 
 process
\begin{equation}
\label{reaction3}
\mu^+~N \rightarrow \mu^+ X^0\pi^{\pm}~N' \rightarrow \mu^+(\Jpsi\pi^+\pi^- )\pi^{\pm} N'\rightarrow \mu^+(\mu^+\mu^- \pi^+\pi^-)\pi^{\pm} N',
\end{equation}
the diagram of which is schematically shown in Fig.~\ref{fig:diag0}.
In order to select such events, we first require a reconstructed vertex in the target region with an incoming beam muon track, three 
outgoing muon tracks (two $\mu^+$, one $\mu^-$) and three outgoing pions ($\pi^+\pi^-\pi^+$ or $\pi^+\pi^-\pi^-$).
Reconstructed particles are identified as muons if they have momentum above 8 GeV/$c$ and have crossed more than 15 radiation lengths of material.
 The muon identification efficiency for such energetic particles is higher than 90\%.
 Other charged particles are assumed to be pions.
  Since the dimuon mass resolution of the setup for the $\Jpsi$ peak is about 50 MeV/$c^2$ \cite{Zc},
 candidates for $\Jpsi$ decaying into a pair of oppositely charged muons are accepted if their reconstructed mass  lies in the range  from 
 3.02~GeV/$c^2$ to 3.18~GeV/$c^2$. 
 With two $\mu^+$ in a given event, we may reconstruct two $\Jpsi$ candidates  in the 
 $\mu^+\mu^-$ final state, in which case the event is rejected ($\sim$3\% of events).
 The nominal $J/\psi$ mass \cite{pdg} is assigned to accepted dimuons.
 In order to select exclusive production in process (\ref{reaction3}), we 
 require $\sum E$ to match the energy $E_\text{beam}$ of the beam particle, except for a small  recoil energy to the target. Here, $\sum E$ is 
 the sum of energies of the scattered muon, of the $\Jpsi$, and of the three pions in the final state.
 Since at COMPASS the experimental resolution for $\Delta E=\sum E-E_\text{beam}$ is about 2~$\GeV$, we require $|\Delta E|<$ 4~$\GeV$
 in order to select exclusive production of the $\Jpsi 3\pi$ final state.
 The total number of selected  exclusive $\mu^+\Jpsi 2\pi^+\pi^-$ and $\mu^+\Jpsi \pi^+2\pi^-$ events is 72 and 49, respectively. 
 The ratio (72/49) corresponds approximately to the ratio of the average numbers of protons and neutrons in the target material that is $\sim1.3$.

Figure \ref{fig:allspectra}(a) shows the mass spectrum for the $\Jpsi \pi^+\pi^-$ subsystem in reaction (\ref{reaction3})  from threshold to 
5~GeV/$c^2$  after the aforementioned selection criteria were applied. As there are two equally charged  pions per event, this mass 
spectrum contains  contributions from the two possible $\pi^+\pi^-$ combinations. The mass spectrum exhibits two peak structures below 
4~GeV/$c^2$, 
with positions and widths that are compatible with the production and decay of $\psi(2S)$ and $X(3872)$. However, for reasons that will be described below, we prefer to name the particle corresponding to the second peak observed for the reaction (\ref{reaction3}) as $\widetilde{X}(3872)$.
 We determine the resonance parameters
 by a maximum likelihood fit to the mass spectrum from threshold to 5~GeV/$c^2$, using a sum of two Gaussian functions for 
 the  two signal peaks
 and the background term
\begin{equation}
B(M)=c_1(M-m_0)^{c_2}e^{-c_3M},
\label{FIT}
\end{equation}
where $M=M_{J/\psi\pi^+\pi^-}$ and $m_0=M_{\Jpsi}+2m_{\pi}$. We ignore possible contributions from other states like $\psi(3770)$, $\psi(4040)$, $\psi(4160)$, 
$X(4260)$, $X(4360)$ and $X(4660)$ since their branching fractions into $\Jpsi \pi\pi$ are too small \cite{pdg} to significantly impact the shape of the 
observed mass distribution.
 The fit function has eight free parameters: the resonance mass and the number of events in each mass peak, the same width 
 $\sigma_{M}$ for both peaks and the parameters $c_1$, $c_2$, $c_3$ describing the background shape. The yields for $\psi(2S)$ and 
 $\widetilde{X}(3872)$ are determined to be $N_{\psi(2S)}=24.2\pm6.5$ and $N_{\widetilde{X}(3872)}=13.2\pm5.2$ events, and their masses are $M_{\psi(2S)}
 =3683.7\pm6.5$~MeV/$c^{2}$ and $M_{\widetilde{X}(3872)}=3860.4\pm10.0$ MeV/$c^{2}$, respectively. The estimated mass values are consistent 
 with the world average values for $\psi(2S)$ and $X(3872)$ \cite{pdg}. 
 The fit yields $\sigma_{M}=22.8\pm6.9$~MeV/$c^{2}$ for the width. As this value is dominated by the experimental resolution, it appears 
 sufficient to use the same width parameter for each Gaussian.
 In order to estimate the statistical significance of the observed signals, the background function $B(M)$ in  Eq.~(\ref{FIT}) was fitted to the mass 
 spectrum shown in Fig. \ref{fig:allspectra}(a) in the region below 5~GeV/$c^2$, excluding the signal range from 3.62~GeV/$c^2$ to 
 3.90~GeV/$c^2$. The probability $p(M)$ to find a number of events equal or larger than observed in the 
 mass window $M\pm\Delta M$, where $\Delta M = $30 MeV/$c^2$, due to a statistical fluctuation, is shown in Fig. \ref{fig:allspectra}(b). In order to calculate $p(M)$ we 
 assume a Poissonian background with the mean value
\begin{equation}
 \bar{N}(M)=\int_{M-\Delta M}^{M+\Delta M} B(M')dM'.
\label{mean}
\end{equation}
  \ifthenelse{\equal{\PRLSTYLE}{yes}}{\begin{figure}}{\begin{figure}[h]}
 \begin{center}
   \includegraphics[width=250px]{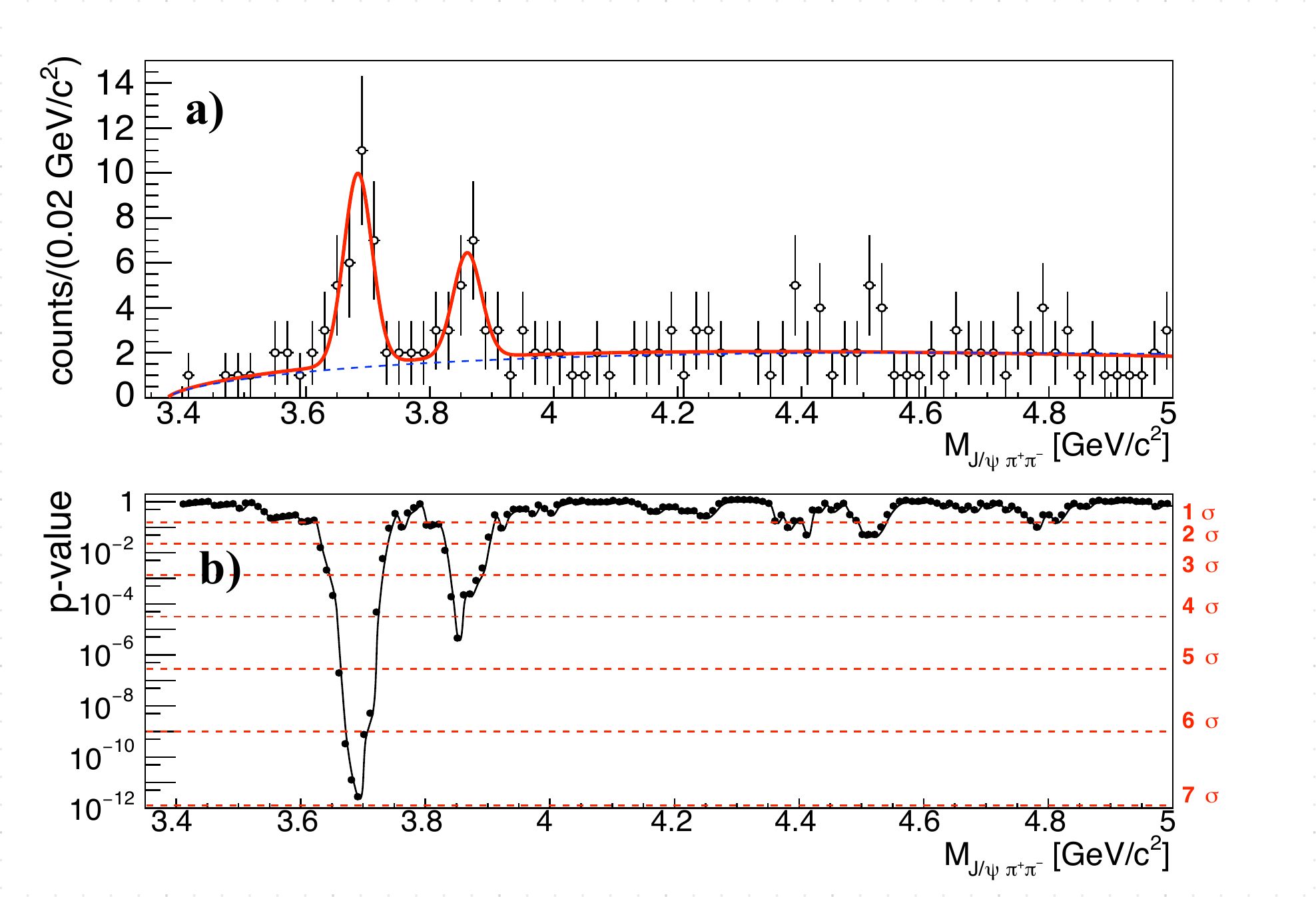}
   \end{center}
  \caption{\label{fig:allspectra}
   (a) The $\Jpsi\pi^+\pi^-$ invariant mass distribution for the $\Jpsi\pi^+\pi^-\pi^{\pm}$ final state (two entries per event) for exclusive  
   events ($|\Delta E|<4$~$\GeV$). The fitted curve is shown in red.  The blue dashed line shows  a fit of the background contribution 
   [Eq. (\ref{FIT})] to the data excluding the signal range. (b) The probability to obtain the observed or a larger number of events due to 
    a statistical fluctuation of the Poissonian background with a mean value described by Eq. (\ref{mean}).}
\end{figure}
The statistical significance for $\psi(2S)$ and $\widetilde{X}(3872)$, expressed in terms of the Gaussian standard deviation, is 6.9$\sigma$ and
 4.5$\sigma$, respectively. A possible contribution of systematic effects is not taken into account here and will be discussed later.
We have repeated the fit keeping the mass separation of the two Gaussians fixed to the 
 mass difference between $\psi(2S)$ and $X(3872)$ from Ref. \cite{pdg},
 which did not significantly alter neither the mass value for the $\psi(2S)$ nor the number
of observed events for either state: $M_{\psi(2S)}=3680.9\pm5.7$ MeV/$c^{2}$, 
$N_{\psi(2S)}=24.9\pm5.7$ and $N_{\widetilde{X}(3872)}=13.6\pm4.8$ events. 

In order to select a non-exclusive data sample for process (\ref{reaction3}), we require a larger missing energy, i.e. 
$-12$~$\GeV<\Delta E<-4$~$\GeV$. The resulting invariant mass distribution is shown in Fig. \ref{fig:allspectra2}(a). Except for $\psi(2S)$, 
we observe no statistically significant signal of charmonium(-like) production.

In parallel to reaction (\ref{reaction3}), we investigate the reaction with neutral exchange,
\begin{equation}
\label{reaction44}
\mu^+~N \rightarrow \mu^+ X^0 N' \rightarrow \mu^+(\Jpsi\pi^+\pi^- ) N'\rightarrow \mu^+(\mu^+\mu^- \pi^+\pi^-) N',
\end{equation}
by requiring in the final state only two charged pions with opposite charge. Hence the schematic representation of reaction
 (\ref{reaction44}) is similar to the one shown in Fig. \ref{fig:diag0}, but without the bachelor pion. The invariant mass distribution for the 
 exclusive $\Jpsi\pi^+\pi^-$ final state is shown in Fig. \ref{fig:diag}.
 \ifthenelse{\equal{\PRLSTYLE}{yes}}{\begin{figure}}{\begin{figure}[h]}
 \begin{center}
   \includegraphics[width=220px]{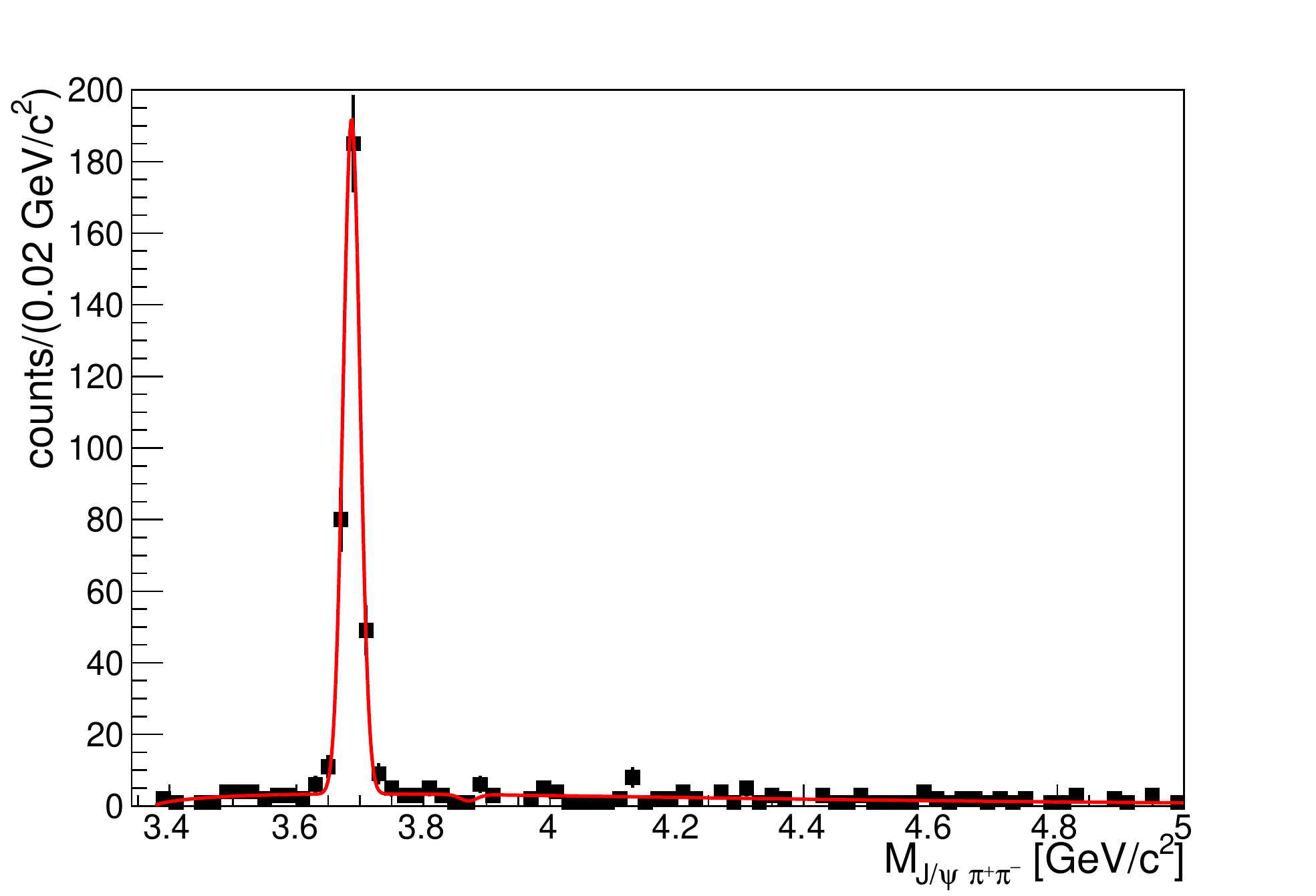}
   \end{center}
     \caption{\label{fig:diag}
 The $\Jpsi\pi^+\pi^-$ invariant mass distribution for the exclusive $\Jpsi\pi^+\pi^-$ final state from reaction (\ref{reaction44}).   }
\end{figure}
  The parameters of the $\psi(2S)$ peak are determined from a fit using the model described
 above with the mass of  the $X(3872)$ Gaussian fixed to the nominal value of the $X(3872)$ mass. They are $N_{\psi(2S)}=314\pm18$, $M_{\psi(2S)}=3687.1\pm0.8$ 
 MeV/$c^{2}$ and $\sigma_{M}=13.3\pm0.7$ MeV/$c^{2}$. 
  The $X(3872)$ yield obtained from the fit is $-2.9\pm2.5$ events, i.e. no statistically significant evidence for an $X(3872)$ signal  was
   found in reaction (\ref{reaction44}). A  statistical simulation was used to determine the upper limit for $N_{X(3872)}$. 
Samples were generated according to the fit results for the $\psi(2S)$ peak and the background continuum, while the strength of the 
$X(3872)$ Gaussian signal was varied. The upper limit  
$N^{UL}_{X(3872)}$ for the number of  events $N_{X(3872)}$, which is required to obtain the result of $-$2.9 events or lower, is 0.9 
events at a confidence level of 90\%. Similar studies were performed for the exclusive reaction with the final state $\mu^+\Jpsi 2\pi^+2\pi^-N'$. It was found that the mass spectrum of the $\Jpsi\pi^+\pi^-$ subsystem does not exhibit any glimpse of  $X(3872)$.

In order to investigate the origins of $\widetilde{X}(3872)$ and $\psi(2S)$ in reaction (\ref{reaction3}), we add the bachelor pion to both states to 
determine the invariant masses of the $\widetilde{X}(3872)\pi^{\pm}$ and $\psi(2S)\pi^{\pm}$ systems. For this study, we consider only the two 
narrow mass regions of $\pm30$ MeV/$c^2$ around the estimated mass values of  $\widetilde{X}(3872)$ and $\psi(2S)$.
The fraction of background events in the samples is 40\% and 25\%, respectively.
 Although no significant structure can be seen in the mass distribution shown in Fig. \ref{fig:kin12}(a), some enhancement of 
 $\psi(2S)\pi^{\pm}$ events may be spotted at masses of about 4~$\GeV/c^{2}$, where the $Z_{c}^{\pm}$(4020) charmonium-like state was 
 observed by BESIII \cite{BES3_Zc1, BES3_Zc4, BES3_Zc2,BES3_Zc3}. Figure \ref{fig:kin12}(b) shows distributions for the
  missing mass, defined as $M_{\text{miss}}^2=( P_{\mu}+P_N-P_{\mu'}-P_{X^{0}})^2$, for reactions (\ref{reaction44}) and 
  (\ref{reaction3}). 
 Note that according to this definition, the bachelor pion contributes to the missing mass of reaction (\ref{reaction3}).
 The mean value of the missing mass for $\psi(2S)$ produced in reaction (\ref{reaction44}) is about 1.4~$\GeV/c^{2}$. When $\psi(2S)$
  and $\widetilde{X}(3872)$ are produced together with a bachelor pion in reaction (\ref{reaction3}), the mean value for the missing mass is 2.7~$\GeV/c^2$ and 4.3~$\GeV/c^2$, respectively. 
 The apparent difference that can be seen between the missing mass distributions for $\psi(2S)$ and $\widetilde{X}(3872)$ produced in reaction 
 (\ref{reaction3}) may indicate different production mechanisms.
The $\Jpsi\pi^+\pi^-$ invariant mass distribution for exclusive $\Jpsi\pi^+\pi^-\pi^{\pm}N'$ events from reaction (\ref{reaction3}) 
 using the additional requirement $M_{\text{miss}}>3$~$\GeV/c^2$ is shown in Fig. \ref{fig:allspectra2} (b). By this requirement the $\psi(2S)$ peak and the background continuum are reduced  in respect to the $\widetilde{X}(3872)$ signal while the statistical significance of the latter decreases to 4$\sigma$.

  \ifthenelse{\equal{\PRLSTYLE}{yes}}{\begin{figure}}{\begin{figure}[h]}
 \begin{center}
   \includegraphics[width=250px]{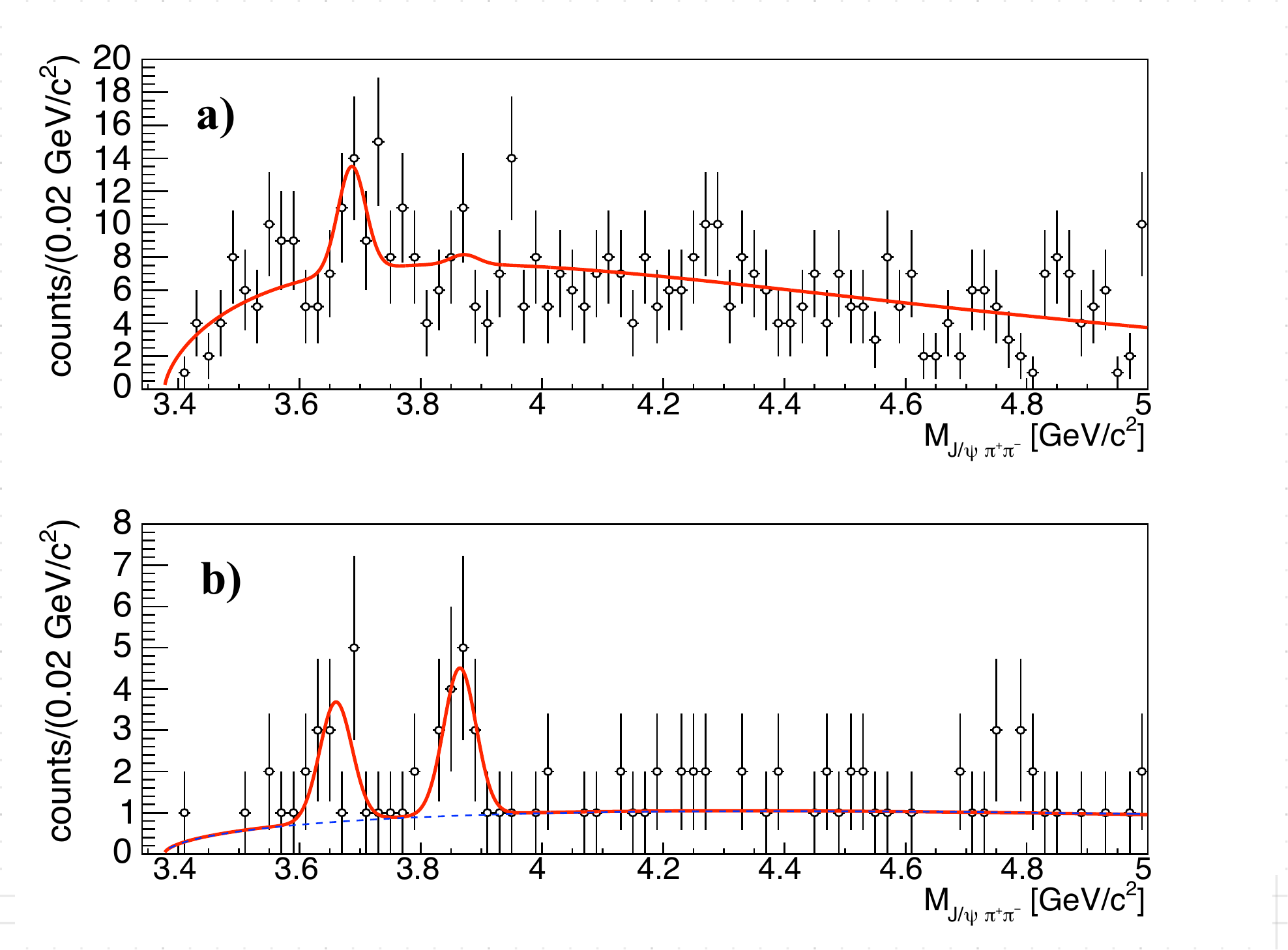}
   \end{center}
  \caption{\label{fig:allspectra2}
 (a) The $\Jpsi\pi^+\pi^-$ invariant mass distributions for the $\Jpsi\pi^+\pi^-\pi^{\pm}$ final state (two entries per event) for non-exclusive  events ($-12$~$\GeV<\Delta E<-4$~$\GeV$) and (b) for exclusive events ($-4$~$\GeV<\Delta E<4$~$\GeV$) with missing mass $M_{\text{miss}}$ above 3~GeV/$c^2$ (see text for the definition of $M_{\text{miss}})$.
 }
\end{figure}

 \ifthenelse{\equal{\PRLSTYLE}{yes}}{\begin{figure}}{\begin{figure}[h]}
 \begin{center}
   \includegraphics[width=220px]{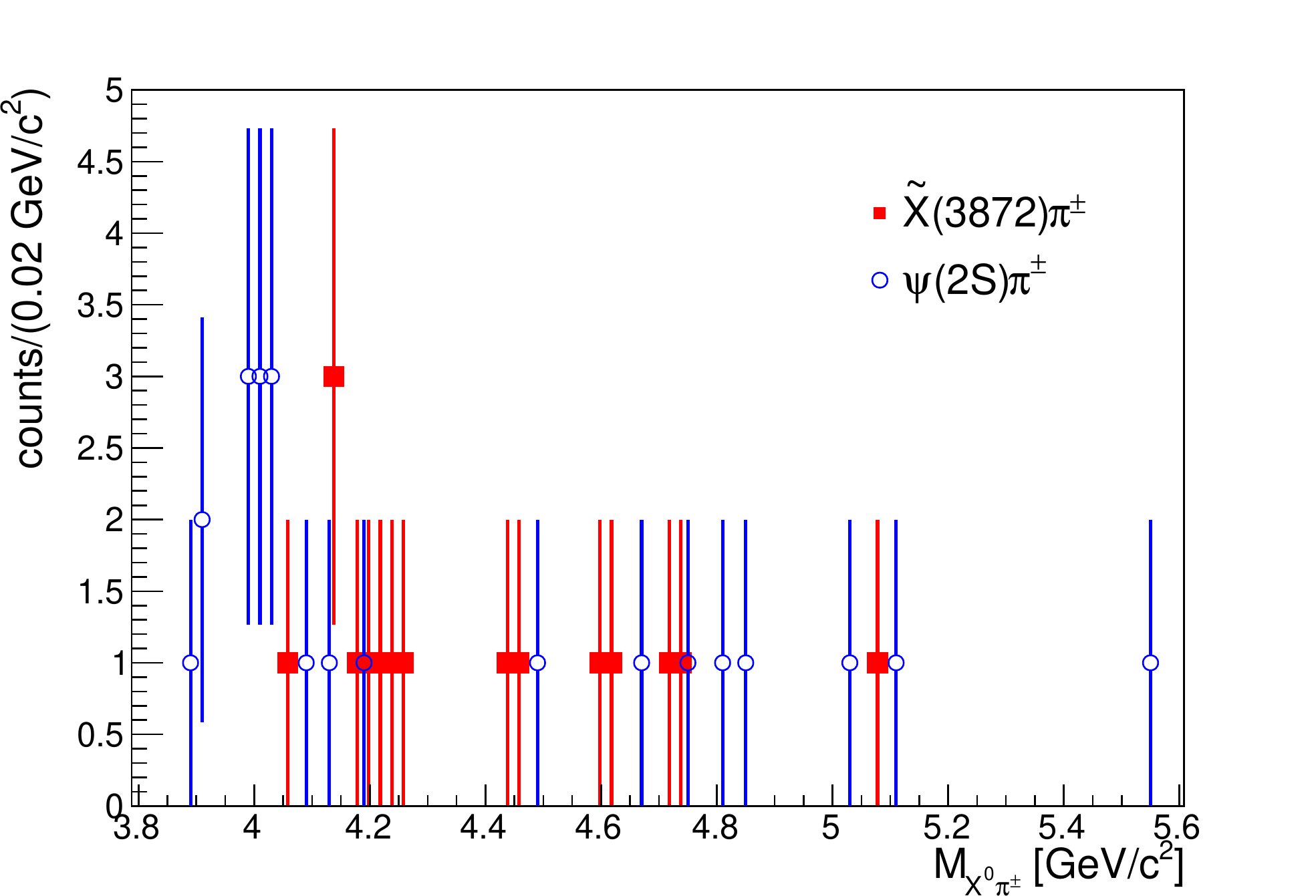}
  \includegraphics[width=220px]{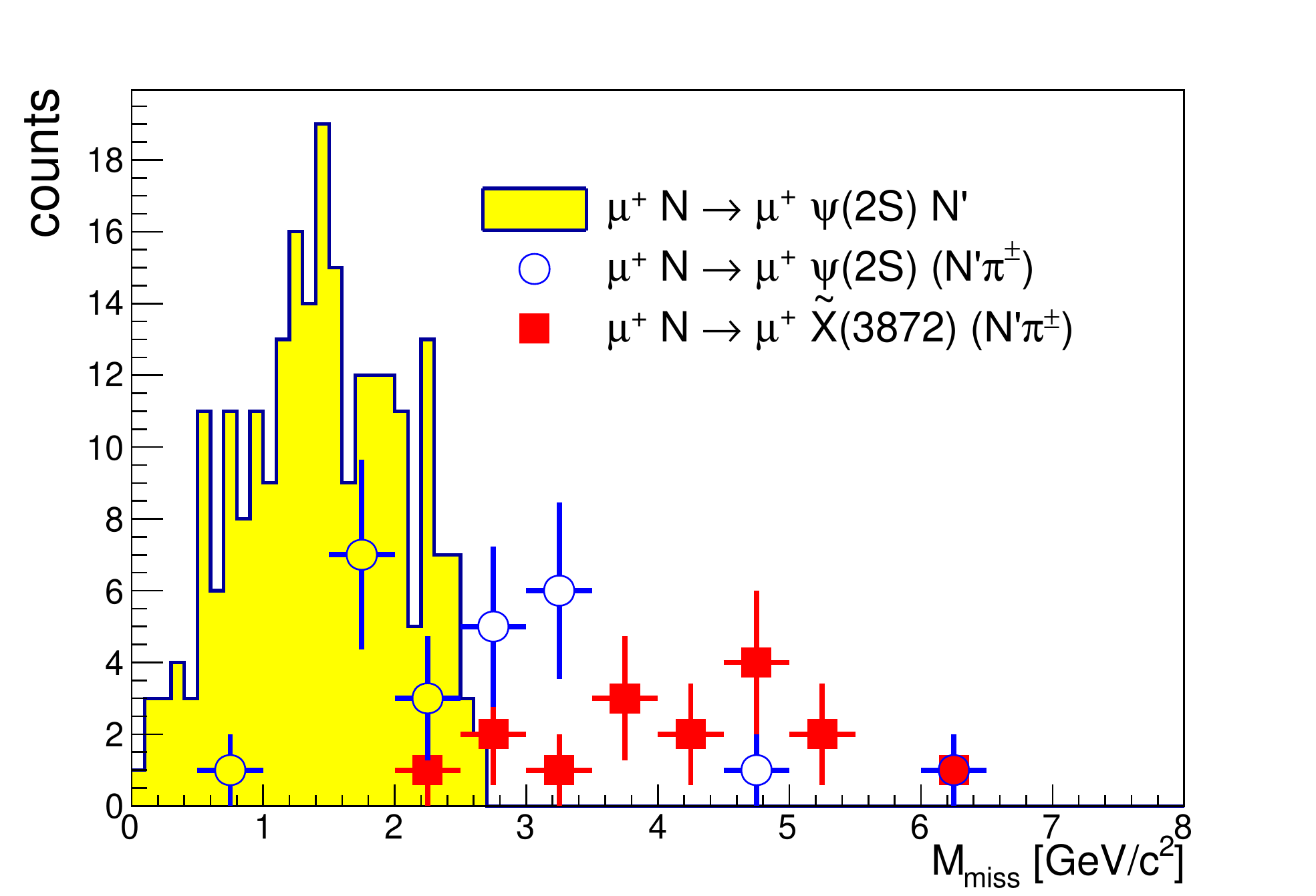}\\
  (a)\hspace{0.4\textwidth}(b)$\quad$ 
   \end{center}
  \caption{\label{fig:kin12}
(a) Invariant mass spectra for $\widetilde{X}(3872)\pi^{\pm}$ (red) and $\psi(2S)\pi^{\pm}$ (blue) of reaction (\ref{reaction3}). (b) Missing mass 
distributions for the exclusive reactions (\ref{reaction3}) and  (\ref{reaction44}). The yellow histogram shows the  events in the range 
$\pm30$~MeV/$c^{2}$ around the $\psi(2S)$ peak of reaction (\ref{reaction44}). Blue circles and red squares 
 show the events in the range $\pm30$~MeV/$c^{2}$ around the $\psi(2S)$ and $\widetilde{X}(3872)$ peaks of reaction (\ref{reaction3}). 
  }
\end{figure}

 \ifthenelse{\equal{\PRLSTYLE}{yes}}{\begin{figure}}{\begin{figure}[h]}
 \begin{center}
   \includegraphics[width=220px]{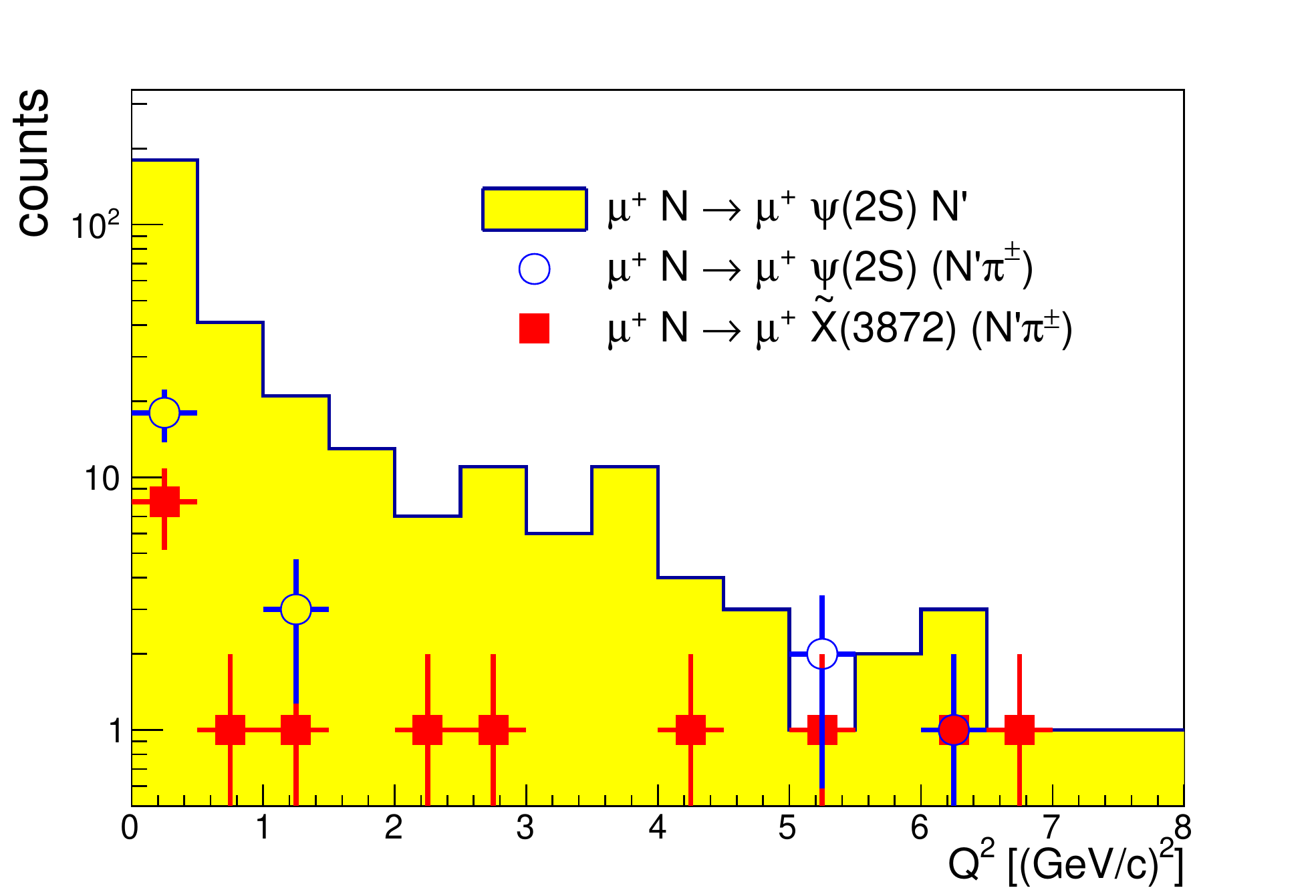}
  \includegraphics[width=220px]{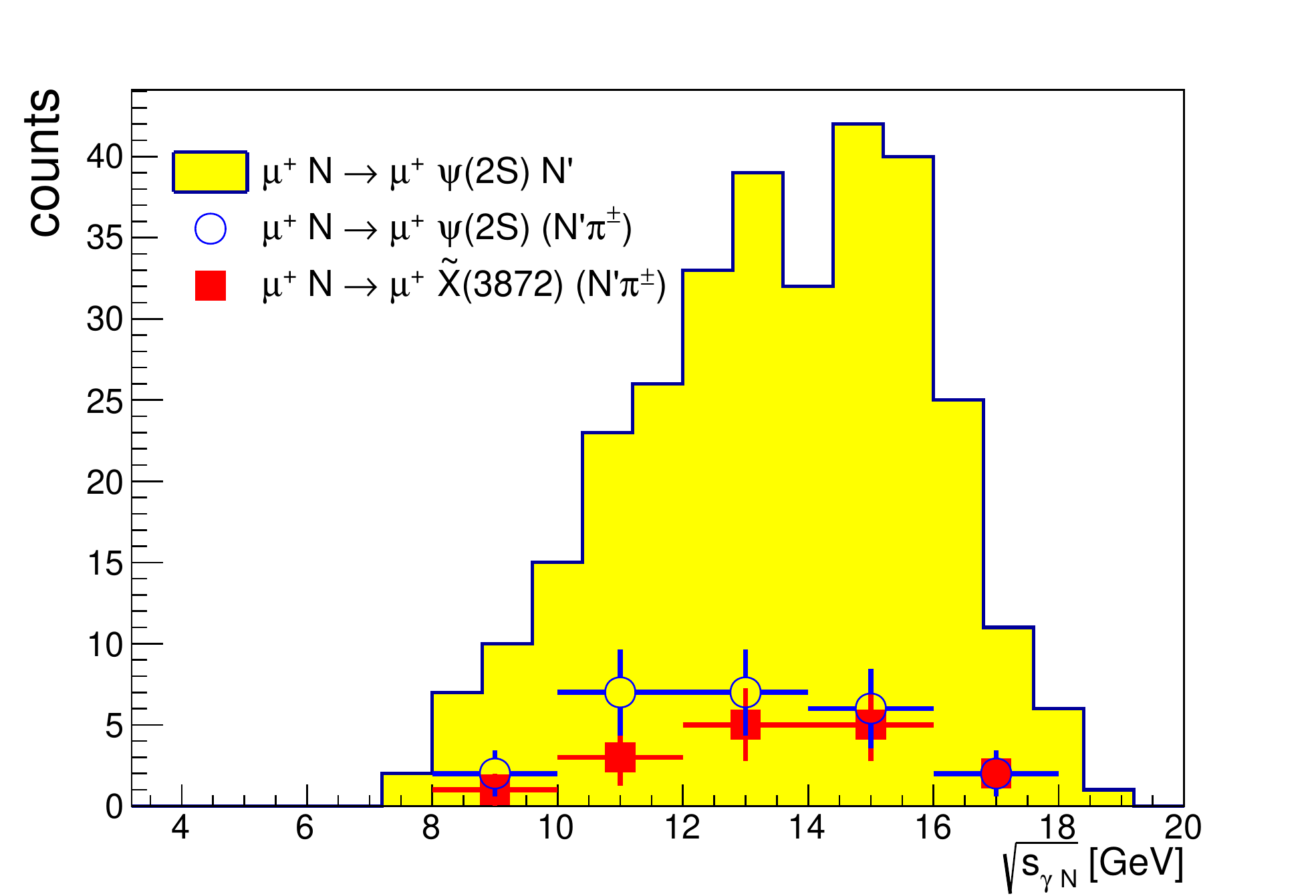}\\
  (a)\hspace{0.4\textwidth}(b)$\quad$ 
   \end{center}
  \caption{\label{fig:kin}
 Kinematic distributions for $Q^2$ (a) and $\sqrt{s_{\gamma N}}$ (b) for reactions (\ref{reaction3}) and  (\ref{reaction44}). 
  The yellow histograms correspond to the  events in the range $\pm30$ MeV/$c^{2}$ around the $\psi(2S)$ peak of reaction 
  (\ref{reaction44}). Blue circles and red squares show the events in the range $\pm30$ MeV/$c^{2}$ around the $\psi(2S)$ and
   $\widetilde{X}(3872)$ peaks of reaction (\ref{reaction3}). }  
\end{figure}

 Reactions (\ref{reaction3}) and (\ref{reaction44}) are characterized by two kinematic variables: the negative squared four-momentum 
 transfer $Q^2 = -( P_{\mu}-P_{\mu'})^2$ and the centre-of-mass (CM) energy of the virtual-photon -- nucleon system,
  $\sqrt{s_{\gamma N}}$. The distributions of these two variables are shown in Figs. \ref{fig:kin}(a) and \ref{fig:kin}(b). Most 
  events occur at small values of $Q^2$. The CM energy is distributed between 8~$\GeV$ and 18~$\GeV$, while the kinematic limit for beam momenta of 160~GeV/$c$  and 200~GeV/$c$ is 17.3~$\GeV$ and 19.4~$\GeV$, respectively.
 We tested the hypothesis that the observed $\widetilde{X}(3872)$ peak is an artificial structure appearing in the reaction 
 $\gamma^*~N \rightarrow  \psi(2S) N^{*} \rightarrow (\Jpsi \pi^+\pi^-) (N' \pi^{\pm})$, where one mixed up the pion from 
 $\psi(2S)$ decay with the pion from $N^{*}$ decay in the reconstruction of the $J/\psi\pi^+\pi^-$ mass. The results of a toy 
 Monte-Carlo simulation disfavour this hypothesis.

 The mass spectrum of the two pions resulting from the decay of the $X(3872)$ was precisely studied, \mbox{e.g.} by the Belle
  \cite{Choi:2011fc}, CDF \cite{Abulencia:2005zc},
  CMS \cite{Chatrchyan:2013cld} and ATLAS \cite{Aaboud:2016vzw} collaborations.  They found a preference for high 
  $\pi^+\pi^-$masses and a 
 dominance of  the $X(3872)\rightarrow \Jpsi\rho^{0}$ decay mode. The measured two-pion mass spectra for events produced in 
 reaction (\ref{reaction3}) 
 within a $\pm$30 MeV/$c^2$ mass window around the $\psi(2S)$ (blue) and the $\widetilde{X}(3872)$ (red) are shown in Fig. \ref{fig:kin21}(a). 
 The result for $\psi(2S)$ is in  a good agreement with former observations, while the shape of the $\pi\pi$ mass distribution for $\widetilde{X}(3872)$ 
 looks  very different from the well-known results for $X(3872)$. The comparison of  the two-pion mass distributions from $\widetilde{X}(3872)$ decay 
 obtained by COMPASS and from $X(3872)$ decay  obtained by ATLAS \cite{Aaboud:2016vzw} (the ATLAS result is taken as a typical high-precision example) is presented in Fig. \ref{fig:kin21}(b). The cut $M_{\text{miss}}>$  3~GeV/$c^2$ is applied for Fig.~\ref{fig:kin21}(b) to reduce underlying background contribution in the $\widetilde{X}(3872)$ sample.
 Our studies show that the observed 
 difference cannot be explained by acceptance effects. Within statistical uncertainties, the shape of the COMPASS $\pi\pi$ mass 
 distribution is in agreement with a three-body phase-space decay and with the expectation for a state with quantum numbers $J^{PC} = 1^{+-}$ 
 \cite{spectrum}, while the quantum numbers previously determined for the $X(3872)$ are $1^{++}$.  A possible distortion of the two-pion mass spectrum by non-resonant background under the peak was estimated using
  the sPlot procedure \cite{Splot} and was found to be unlikely for reaction (\ref{reaction3}).
 The statistical significance of the disagreement between the observed two-pion mass spectrum and the expected one from the known decay $X(3872)\rightarrow \Jpsi\rho^{0}$ was estimated using the maximum likelihood approach and was found to be between 4.7$\sigma$ and 7.3$\sigma$ depending on the treatment of the residual background under the $\widetilde{X}(3872)$ peak.
 We 
  investigated the possibility to obtain the observed two-pion spectrum from the decay $X(3872)\rightarrow \Jpsi\omega \rightarrow \Jpsi\pi^+\pi^-\pi^0$ where the $\pi^0$ has been lost, and excluded it. A possibility to have visible contribution  from the $\chi_{c0,1,2}\rightarrow \Jpsi\gamma$ decay, followed by the photon conversion into $e^+e^-$ misidentified as $\pi^+\pi^-$, was also investigated and 
excluded. 
A possible interpretation of the observed $\widetilde{X}(3872)$  signal is that it is not the well-known $X(3872)$ but a new charmonium state with similar mass. This would be in agreement with the tetraquark model of Refs. \cite{Maiani:2004vq, Maiani:2014aja} which predicts a neutral  partner of  $X(3872)$ that has a similar mass, negative $C$-parity, and decays
 into $J/\psi\sigma$.

 \ifthenelse{\equal{\PRLSTYLE}{yes}}{\begin{figure}}{\begin{figure}[h]}
 \begin{center}
  \includegraphics[width=220px]{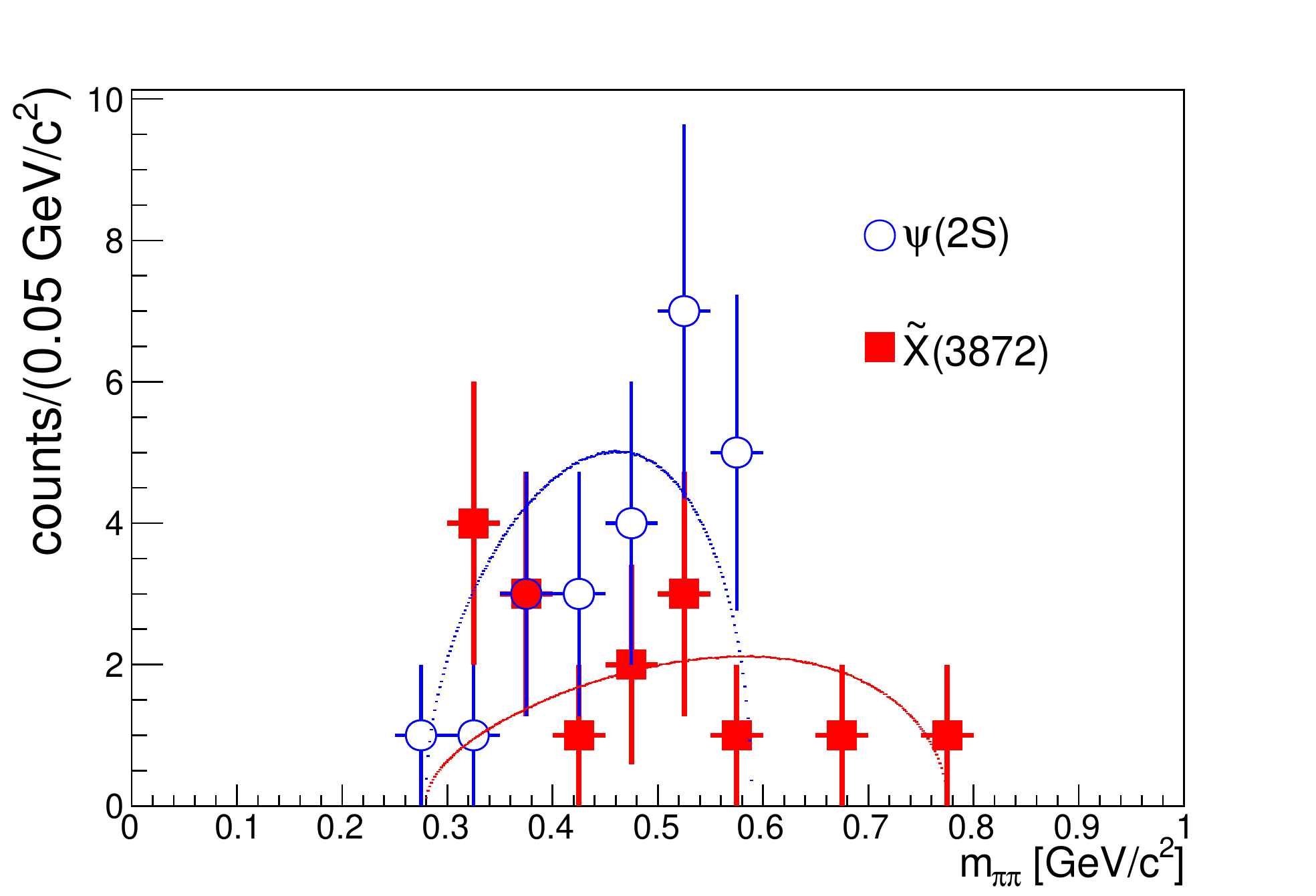}
\includegraphics[width=220px]{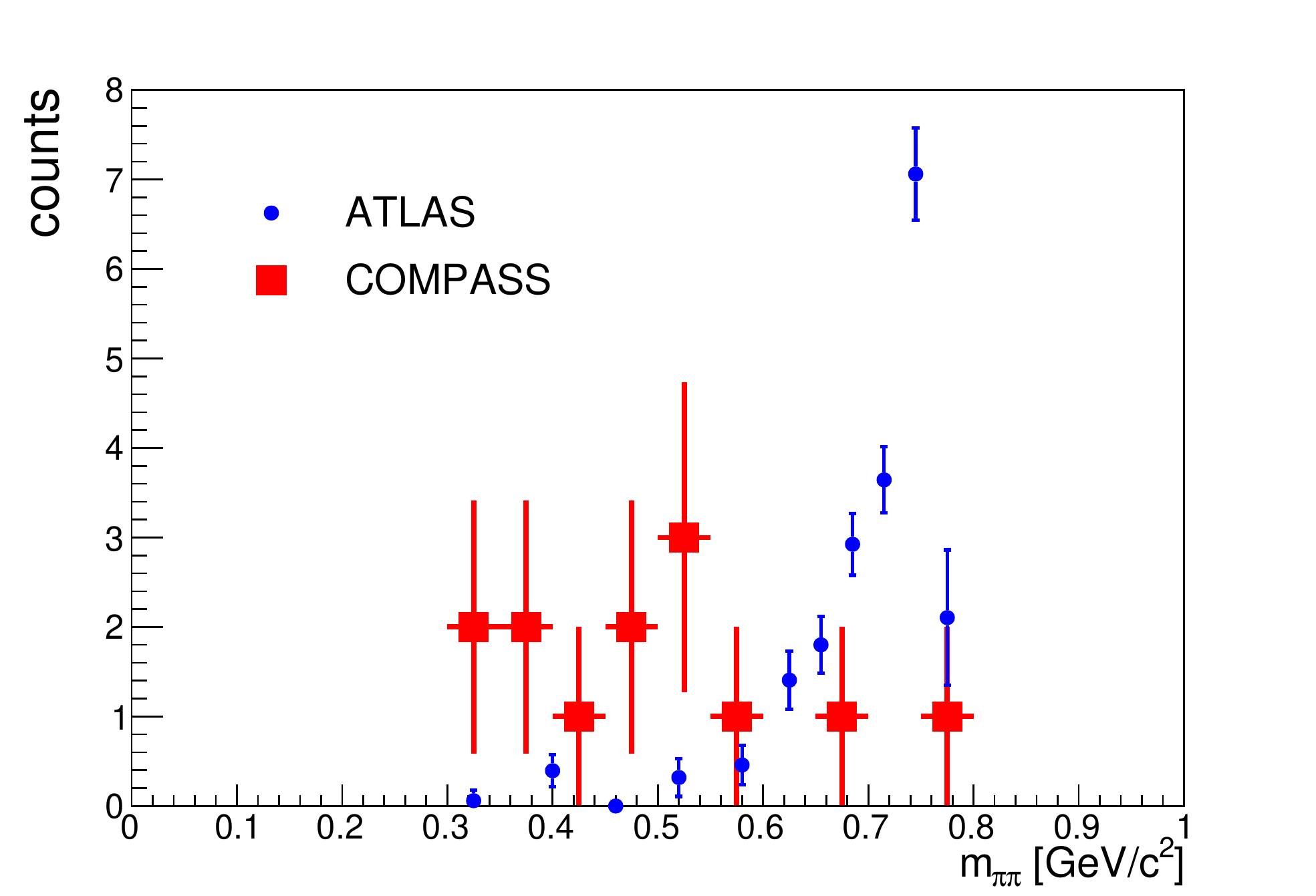}\\
  (a)\hspace{0.4\textwidth}(b)$\quad$ 
   \end{center}
  \caption{\label{fig:kin21}
 (a) Invariant mass spectra for the $\pi^+\pi^-$ subsystem from the decay of $\widetilde{X}(3872)$ (red squares) and $\psi(2S)$ (blue circles) 
 produced in reaction (\ref{reaction3}).  The corresponding distributions for three-body phase-space decays are shown by the curves. 
 (b) Invariant mass spectra for the $\pi^+\pi^-$ subsystem from the decay of  $\widetilde{X}(3872)$ measured by COMPASS 
  with the applied cut $M_{\text{miss}}>$  3~GeV/$c^2$ (red squares) and from the decay of $X(3872)$ observed by 
 ATLAS \cite{Aaboud:2016vzw} (blue points). Both distributions are normalised to the same area.}
\end{figure}

In order to estimate the Breit-Wigner width of the $\widetilde{X}(3872)$ state the fitting procedure
for the $\Jpsi\pi^+\pi^-$ invariant mass distribution shown in Fig. \ref{fig:allspectra}(a) was redone. A Gaussian shape was used to fit the $\psi(2S)$ peak while the convolution of a Gaussian distribution of the same width as for $\psi(2S)$ and a Breit-Wigner function having the same mass as the Gaussian one was used for $\widetilde{X}(3872)$. The obtained result for the width of $\widetilde{X}(3872)$ is the upper limit $\Gamma_{\widetilde{X}(3872)}<$ 51 MeV/$c^2$ CL = 90\%.

The previously mentioned statistical significance of the $\widetilde{X}(3872)$ signal was evaluated without including systematic effects.
As a result of the comprehensive studies of systematic effects, we conclude that the systematic uncertainty related to our choice of the background shape [Eq. (\ref{FIT})] and the fitting range is the dominant one. We estimate this uncertainty to be equivalent to  15\% of the Gaussian uncertainty of the $\bar{N}$ value [Eq. (\ref{mean})]. 
Taking into account this systematic uncertainty by using the frequentist approach proposed in Ref. \cite{Cousins:2007bmb}, the significance of the $\widetilde{X}(3872)$ signal shown in  Fig. \ref{fig:allspectra}(b) is reduced from 4.5$\sigma$ to 4.1$\sigma$. 
We quote the latter value as the estimate of significance of the $\widetilde{X}(3872)$ signal.

 In order to determine the cross section of exclusive $\widetilde{X}(3872)$ production in reaction (\ref{reaction3}), we use  the exclusive production of 
 $\Jpsi$ off the target nucleon,
 \begin{equation}
\label{reaction32}
\mu^+~N \rightarrow \mu^+ \Jpsi~N,
\end{equation}
as normalization.
The same data are used and the same selection criteria are applied as for reactions (\ref{reaction3}) and (\ref{reaction44}). The method
 used to determine the cross section for reaction (\ref{reaction3}) relies on the assumption that the fluxes of virtual photons for 
 reactions (\ref{reaction3}) and (\ref{reaction32}) are the same.
  This assumption is supported by the similar shapes of the $Q^2$ and $\sqrt{s_{\gamma N}}$ distributions in both cases.
 We can therefore relate the photo- and leptoproduction cross sections as follows:
\begin{equation}
\label{reaction31}
\frac{\sigma_{\mu~N \rightarrow \mu \widetilde{X}(3872)\pi~N'}}{\sigma_{\mu~N \rightarrow \mu \Jpsi~N}}=\frac{\sigma_{\gamma~N \rightarrow \widetilde{X}(3872)\pi~N'}}{\sigma_{\gamma~N \rightarrow \Jpsi~N}}.
\end{equation}
 The cross section of the reaction $\gamma~N \rightarrow \Jpsi~N$ is known for our range of $\sqrt{s_{\gamma N}}$; it is 
 $14.0\pm1.6$(stat) $\pm2.5$(syst)~nb at $\sqrt{s_{\gamma N}} ~ = $ 13.7~$\GeV$ \cite{NA14}.
 Since this value was obtained  for the production by a real-photon beam, we reduce it by a factor of 0.8 in order to take into account the 
 $Q^{2}$ dependence of the cross section by using the parameterisation of Ref. \cite{ZEUS}  and the average photon virtuality in our samples of about 1~(GeV/$c)^{2}$.
Since the three charged pions appear only in the final state of reaction (\ref{reaction3}), the ratio of acceptances of the two reactions is in 
first approximation equal to the pion acceptance  $a_{\pi}$ cubed. Based on previous COMPASS measurements and Monte Carlo 
simulations, we estimate $a_{\pi}=0.6\pm0.1(\text{syst})$ as average over the geometrical detector acceptance and both target 
configurations.  Thus we set
\begin{equation}
\frac{\sigma_{\gamma N \rightarrow \widetilde{X}(3872)\pi N'}\times \pazocal{B}_{\widetilde{X}(3872)\rightarrow \Jpsi \pi\pi}}{\sigma_{\gamma N \rightarrow \Jpsi N}}=\frac{N_{\widetilde{X}(3872)}}{a_{\pi}^3N_{\Jpsi}},
\end{equation}
where $N_{\widetilde{X}(3872)}$ and $N_{\Jpsi}$ are the respective numbers of observed $\widetilde{X}(3872)$ and $\Jpsi$ events
from exclusive production on quasi-free nucleons. The number $N_{\Jpsi}$ is determined as  $9.6\times 10^3$, with a systematic 
uncertainty of about 10\% due to non-exclusive background in our data sample. The amount of COMPASS data used in this analysis
  is equivalent to about 14~pb$^{-1}$ of the integrated luminosity, when considering a real-photon beam of about 100~$\GeV$ incident energy 
  scattering off free nucleons. Using the normalization procedure described in Ref. \cite{Zc}, we determine the cross section for  the reaction
   $\gamma N \rightarrow \widetilde{X}(3872)\pi^{\pm} N'$ multiplied by the branching fraction for the decay $\widetilde{X}(3872)\rightarrow \Jpsi \pi^+\pi^-$ to be 
\begin{equation}
\sigma_{\gamma N \rightarrow \widetilde{X}(3872)\pi N'}\times \pazocal{B}_{\widetilde{X}(3872)\rightarrow \Jpsi \pi\pi}= 71\pm28(\text{stat})\pm39(\text{syst})~{\mbox{pb}}.
\end{equation}
The statistical uncertainty is given by the uncertainty in the number of $\widetilde{X}(3872)$ signal events, while the main contributions to the 
systematic uncertainty are: (i) 36~pb from the estimation of $a_{\pi}^3$, (ii) 14~pb from the cross section for reaction (\ref{reaction32}), (iii) 7~pb from the 
estimation of $N_{\Jpsi}$. 

Also, an upper limit is determined for the production rate of $X(3872)$ in the reaction $\gamma N \rightarrow X(3872) N$, 
mentioned in Ref. \cite{BingAnLi}, using the same procedure for normalization as described above. The result is 
\begin{equation}
\sigma_{\gamma N \rightarrow X(3872)N'}\times \pazocal{B}_{X(3872)\rightarrow \Jpsi \pi\pi}< 2.9~{\mbox{pb  (CL}=90\%)}.
\end{equation}

 In summary, in our study of the process depicted  in Fig.~\ref{fig:diag0} we observed the muoproduction of the state $\widetilde{X}(3872)$
with a statistical significance of 4.1$\sigma$. The absolute production rate of this state in $\Jpsi\pi^+\pi^-$ mode was also measured. Its mass $M_{\widetilde{X}(3872)}=3860.0\pm10.4$ MeV/$c^2$ and width $\Gamma_{\widetilde{X}(3872)}<$ 51 MeV/$c^2$ CL=90\% and decay mode $\widetilde{X}(3872)\rightarrow \Jpsi \pi\pi$ are consistent with the $X(3872)$. Our observed two-pion mass spectrum shows disagreement with previous experimental results for the $X(3872)$. 
A possible explanation could be that the observed state is the C = $-$1 partner of the $X(3872)$ as predicted by a tetraquark model. The presented results  demonstrate the physics potential of studying exotic charmonium-like states in (virtual) photoproduction. However, an independent confirmation of the nature of the observed $\widetilde{X}(3872)$ signal from high-precision experiments with high-energy virtual or real photons is required.

We gratefully acknowledge the support of the CERN
management and staff as well as the skills and efforts of the
technicians of the collaborating institutions. We are also grateful to Dmitry Dedovich, Simon Eidelman, Christoph Hanhart, Luciano Maiani, Sebastian
 Neubert, Mike Pennington, Eric Swansen and Adam Szczepaniak for fruitful discussions.

\ifthenelse{\equal{\PRLSTYLE}{yes}}{
This work is supported
by ...}{}

\ifthenelse{\equal{\PRLSTYLE}{yes}}{\bibliography{picpol}}
{}
\end{document}